\documentclass[twocolumn,prb,aps,floatfix]{revtex4}% Physical Review B

\usepackage{graphicx}% Include figure files
\usepackage{dcolumn}% Align table columns on decimal point
\usepackage{bm}% bold math

\begin{document}

\title{Effect of Hole Doping on the London Penetration Depth of
Bi$_{2.15}$Sr$_{1.85}$CaCu$_{2}$O$_{8+\delta}$ and
Bi$_{2.1}$Sr$_{1.9}$Ca$_{0.85}$Y$_{0.15}$Cu$_{2}$O$_{8+\delta}$}% Force line breaks with \\

\author{W. Anukool$^{1,2}$, S. Barakat$^1$, C. Panagopoulos$^{1,3,4}$ and J. R. Cooper$^1$}
\affiliation{$^1$Physics Department, University of Cambridge,  Cambridge CB3 0HE,
UK\\$^2$Physics Dept., Chiang Mai University, Thailand, 50200\\$^3$Department of
Physics, University of Crete and FORTH, 71003 Heraklion, Greece,\\$^4$Department of
Physics and Applied Physics, Nanyang Technological University, Singapore}

\date{Submitted 20th October 2008, Revised 10th June 2009}% It is always \today, today,

\begin{abstract}We report measurements of AC susceptibility and hence the in-plane London penetration
 depth on the same samples of Bi$_{2.15}$Sr$_{1.85}$CaCu$_{2}$O$_{8+\delta}$ and
Bi$_{2.1}$Sr$_{1.9}$Ca$_{0.85}$Y$_{0.15}$Cu$_{2}$O$_{8+\delta}$ for many values of the
hole concentration ($p$).  These  support the scenario in which  the pseudogap weakens
the superconducting response only for $p\lesssim 0.19$.
\end{abstract}


 \maketitle

\section{Introduction}
It is generally accepted that there is  a pseudogap (PG)  in the low energy excitation
spectrum of hole-doped cuprate superconductors, with a definite  energy scale ($E_G
\equiv k_BT^*$) that falls as  the hole concentration $p$ is increased. However the
details are controversial and a recent review \cite{HufnerRev} gives equal weight to
three possible scenarios: (A) $T^*(p)$ falls to zero on the over-doped side together
with the superconducting transition temperature $T_c(p)$, (B) $T^*(p)$ falls sharply
to zero for slightly over-doped samples with $p\simeq 0.19$ or (C) $T^*(p)$ is similar
to case (B) but there is no PG in the superconducting state. The $p$-dependences of
the heat capacity \cite{Loram-02},
 and the in-and out-of-plane penetration depths,
$\lambda_{ab}$ and $\lambda_{c}$, of two grain-aligned single layer cuprates
\cite{Panagop-1,Panagop-mono} suggest that at low $T$ the condensation energy
\cite{Loram-02} and the superfluid density \cite{Panagop-1,Panagop-mono} fall abruptly
below $p\simeq0.19$. This seems to  support scenarios of type  B, but the issue is
still being debated \cite{HufnerRev, Anderson,Fischer}. Here we report AC
susceptibility (ACS) data \cite{AnukoolPhD} showing how  the in-plane superconducting
penetration depth $\lambda_{ab}$ of Bi$_{2.15}$Sr$_{1.85}$CaCu$_{2}$O$_{8+\delta}$
(Bi:2212) and Bi$_{2.1}$Sr$_{1.9}$Ca$_{0.85}$Y$_{0.15}$Cu$_{2}$O$_{8+\delta}$
(Bi(Y):2212) changes between $p$ = 0.08 and 0.21. These are  relatively
straightforward measurements, on the same unaligned sample  as a function of $\delta$,
that can easily be checked by other research groups. We believe that they also support
scenarios of type B above. They also give a linear $T$ - dependence of
$1/\lambda_{ab}(T)^2$, over a wide range of $T$, possibly becoming   ``super-linear''
for pure Bi:2212 below 20 $K$.  Combining  our large values of $\lambda_{ab}(0)$ at
low $p$ with recent scanning tunnelling spectroscopy (STS) work  \cite{KohsakaDavis},
suggesting relatively large Fermi arcs for $T_c$ = 20 and 45 $K$, may  also give
important insights into cuprate superconductivity.
\begin{table}
\caption{\label{table01}Annealing conditions, hole concentrations and room temperature
thermopower for Bi:2212 and Bi(Y):2212 samples. For both powder and bulk samples
Eqn.~\ref{p1-eq01} has been used to find $p$  from $T_{c}$ measured by ACS, with
$T_{c}^{max}=87.2\pm 0.3$ $K$ for Bi:2212 and $89.3\pm 0.3$ $K$ for Bi(Y):2212. The
measured values of the room temperature thermoelectric power, S(290) are also shown,
and give consistent values of $p$.}
\begin{ruledtabular}
\begin{tabular}{llccc}
{Quench}&{Anneal conditions}&\multicolumn{2}{c}{$p$ (holes/CuO$_{2}$)}&{S(290)}\\
&&Powder&Bulk&($\mu V/K$)\\ \hline
Bi:2212&&&&\\
Q1&$300^{\circ}C$, 100\%O$_{2}$, 7days&0.207&0.208&-6.32\\
Q2&$400^{\circ}C$, 100\%O$_{2}$, 24h&0.194&0.196&-4.18\\
Q3&$450^{\circ}C$, 100\%O$_{2}$, 23.5h&0.188&0.187&-3.06\\
Q4&$500^{\circ}C$, 100\%O$_{2}$, 24h&0.182&0.181&-1.81\\
Q5&$600^{\circ}C$, 100\%O$_{2}$, 23.5h&0.160&0.162&1.91\\
Q6&$550^{\circ}C$, 80ppmO$_{2}$, 24h&0.154&0.160&3.24\\
Q7&$620^{\circ}C$, 80ppmO$_{2}$, 24h&0.148&0.146&4.98\\
Q9&$580^{\circ}C$, Vacuum, 24h&0.124&0.123&11.25\\
Q13&$650^{\circ}C$, Vacuum, 22h&0.105&0.105&15.06\\ \hline
Bi(Y):2212&&&&\\
Q1&$300^{\circ}C$, 100\%O$_{2}$, 7days&0.182&0.181&-2.2\\
Q3&$350^{\circ}C$, 100\%O$_{2}$, 24h&0.172&0.170&-0.63\\
Q4&$425^{\circ}C$, 100\%O$_{2}$, 24h&0.160&0.160&1.39\\
Q6&$550^{\circ}C$, 100\%O$_{2}$, 24h&0.151&0.148&5.90\\
Q8&$550^{\circ}C$, 10\%O$_{2}$, 30h&0.144&0.140&7.11\\
Q10&$600^{\circ}C$, 80ppmO$_{2}$, 24h&0.121&0.120&14.24\\
Q11&$540^{\circ}C$, Vacuum, 24h&0.098&0.099&25.25\\
Q12&$600^{\circ}C$, Vacuum, 24h&0.088&0.094&28.73\\
Q15&$650^{\circ}C$, Vacuum, 24h&0.082&0.087&33.64\\
\end{tabular}
\end{ruledtabular}
\end{table}

\begin{figure}[hbtp]
\begin{center}
\includegraphics[width=70mm,height=95mm]{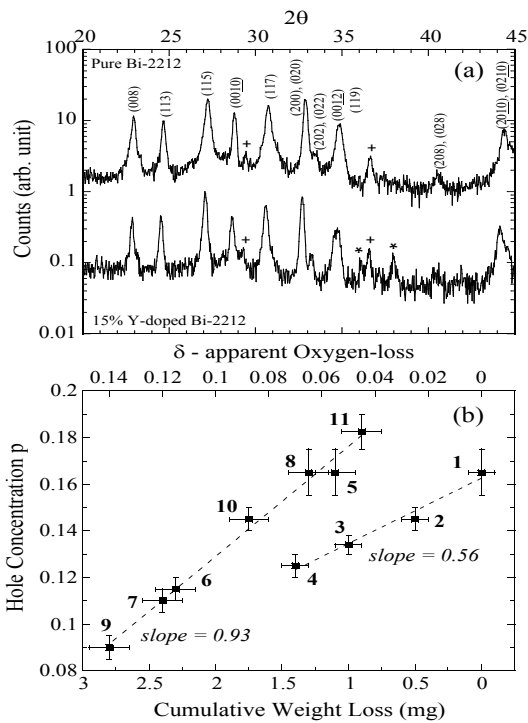}
\caption{(a) powder Cu K$\alpha$  XRD patterns for Bi:2212 and  Bi(Y):2212 on a
logarithmic scale. Impurity lines from Bi$_{2}$O$_{2.5}$ or Bi$_{2}$O$_{3}$ are marked
by (*) and incommensurate satellite reflections \cite{Xrayincomm} by (+). (b)
 $p$  is plotted  \textit{vs.} weight loss - lower scale and
apparent oxygen loss - upper scale. Heat treatment details are given in footnote
\onlinecite{Footnote2}. $p$ values were found from the ACS $T_c$ of a sintered
Bi(Y):2212 bar using Eq. 1 and $T_c^{max}= 89.1 \pm0.3 K$. } \label{ch03-fig01}
\end{center}
\end{figure}

 \section{Experimental Details}
X-ray powder diffraction (XRD) patterns for the two samples in Fig.~\ref{ch03-fig01}a.
show that the  Bi:2212 sample was phase pure to within the noise level of $ 2-3\%$.
Two peaks that are not indexed arise from the incommensurate superstructure
\cite{Xrayincomm}.  They are also present for Bi(Y):2212 but this has two more peaks
from  3-5\% Bi$_{2}$O$_{2.5}$ or Bi$_{2}$O$_{3}$. Examples of how the  oxygen content
of the sintered and powder samples was varied by annealing in flowing gases  and
quenching into liquid nitrogen \cite{Footnote1} are given in Table 1. Sample weights
were measured immediately after quenching. $p$ values were obtained from the $T_c$
values measured by ACS  using  the parabolic law~\cite{Presland-01}:

\begin{equation}\frac{T_{c}}{T_{c}^{max}} = 1-82.6(p-0.16)^{2}
\label{p1-eq01}
\end{equation}

Fig.~\ref{ch03-fig01}b \cite{Barakatproject} shows  $p$
 \textit{vs.}  cumulative weight changes
for 11 annealing treatments on one of the  sintered Bi(Y):2212 samples. After the
initial irreversible changes (steps 1-4), the data are reversible and give a line of
unit slope \cite{Footnote2}. This provides further experimental confirmation
\cite{Presland-01} of the \textit{changes} in $p$ during the quenching treatments,
because oxygen in the Sr-O Bi-O reservoir layer is expected to be O$^{2-}$ and there
are 2 Cu atoms per formula unit. Fig.~\ref{ch03-fig01}b also shows that the
irreversible losses ($\simeq$ 1 $mg$ in 1.1 $g$) occurred continuously on heating from
450 to $550^{\circ}C$ \cite{Footnote2} and do not affect the value of $p$ (since the
$p$ values at steps 1 and 5 are the same). Loss of Bi is the most likely cause since
such a weight loss is equivalent to 0.004 Bi per formula unit and for a valency
Bi$^{4+}$ this would only change $p$ by 0.008 - well within the error bars of points 1
and 5 in Fig.~\ref{ch03-fig01}b.

Fine powders were obtained by gently grinding  50-100 $mg$ of the fully-oxygenated
sintered material with a small pestle and mortar. Two series of ACS experiments on
Bi(Y):2212 powders  gave similar results, data shown in Table 1 and Figs. 2b and 3b
are for the second set.  After the final (13th) quench, grain sizes were determined by
measuring the dimensions of $\sim500$ grains in a scanning electron microscope (SEM)
photograph. Approximately 1/3 of the grains were not circular and for these the
geometric mean of the two radii was used.  For Bi(Y):2212 the grain radius ($r$) at
50\% cumulative volume (CV) was 1.5 $\mu m$, with $r$ = 0.53 and 2.20 $\mu m$ at 10
and 90\% CV respectively. For pure Bi:2212 the powders were sedimented in acetone to
remove large particles, giving $r = 0.65$, 0.31 and 1.01 $\mu m$ for the 50, 10 and
90\% CV points respectively.

\begin{figure}[hbtp]
\begin{center}
\includegraphics[width=70mm,height=100mm]{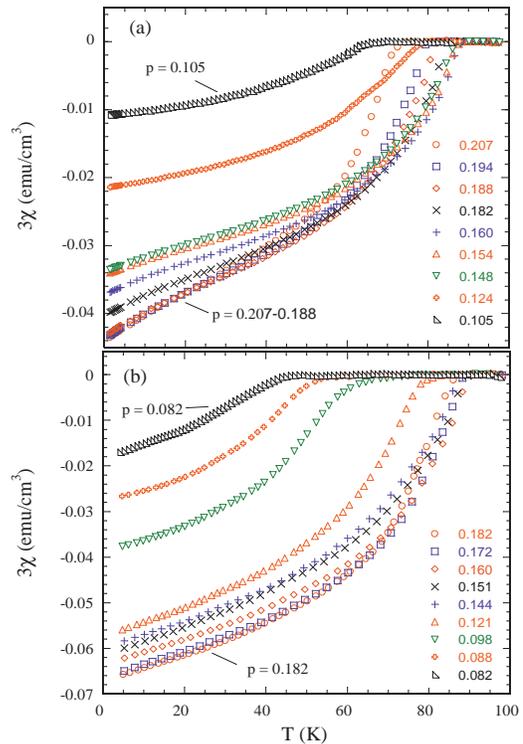} \caption{Color on-line: volume
susceptibility $\chi(T)$ vs. $T$ for (a) pure Bi:2212 and (b) 15\% Y-doped Bi:2212.
The $p$ values determined from $T_{c}$ are also shown.} \label{ch05-fig02}
\end{center}
\end{figure}

ACS measurements  were made using standard techniques \cite{Footnote3}. In the limit
where $\lambda_{c}$ is much larger than both $\lambda_{ab}$ and the grain size,  we
can obtain $\lambda_{ab}$ since only the component of $H$ along the crystalline $c$ -
axis will give rise to detectable diamagnetism. In this case the AC signal from
randomly oriented grains is  1/3 of that for aligned grains with $H\parallel c$,
because the average of $\cos^{2}\theta$ over a sphere is 1/3. Therefore we simply
multiply the observed AC signal by 3 and  apply our usual analysis for aligned
grains~\cite{Porch-01,Panagop-mono} with $H\parallel c$.  The ACS data in
Figs.~\ref{ch05-fig02}a and~\ref{ch05-fig02}b are in emu/cm$^3$. Multiplying by
$8\pi/3$ gives $m/m_{max}$ where the magnetic moment $m$ is reduced from $m_{max}$ for
a perfectly diamagnetic sphere because of the finite value of $\lambda_{ab}$. Allowing
for a finite value of $\lambda_{c}$ has negligible effect  \cite{Footnote4}.

The ACS signals in Fig.~\ref{ch05-fig02} were extrapolated linearly to zero to find
$T_{c}$ giving $T_{c}^{max}=87.2\pm 0.3$ $K$ for Bi:2212 and $89.3\pm 0.3$ $K$ for
Bi(Y):2212, and $p$ values  were then found  using Eq.~\ref{p1-eq01}. We note that the
lower values of $p$ in Table 1 were obtained by vacuum annealing which gives less
control than using flowing gases.  We believe that these samples are nevertheless
uniformly doped for following reasons. (i) the samples were periodically
re-oxygenated, e.g. between quenches Q9 and Q13 and Q12 and Q15 in Table 1, and the
results showed that the vacuum anneals  did not cause any irreversible changes.  (ii)
At $600^{\circ}C$ oxygen  diffusion  in Bi:2212 \cite{Bensemandiff} is so fast that
the oxygen content in a  10 $\mu$m diameter grain will attain uniformity in 1 second.
(iii) The hole concentrations in Table 1   obtained from $T_c$ for powders and
sintered bars are very similar and agree with $p$ values determined from the room
temperature thermoelectric power using the results of Ref. \onlinecite{OCT}. Such good
agreement would not be obtained if the oxygen content of the powder samples were
substantially non-uniform.

  The raw ACS data in Fig.~\ref{ch05-fig02} clearly
show that the signal, i. e. $\lambda_{ab}$ , only changes strongly with $p$ on the
under-doped side for $p \leq 0.19$ and that
 for the Bi:2212 sample it saturates for $p \geq 0.188$.  Also the well-defined onsets in the
 ACS signals
 at $T_{c}$ are consistent with the finite size scaling analysis of specific heat
data~\cite{Loram-scaling}, that ruled out gross inhomogeneity at any doping level. For
Bi:2212 there is an increase in the slope of the diamagnetism below  20 $K$ for
$0.15<p<0.21$, but this is not visible for the Y-doped samples over the range of $p$
values studied, nor for lightly Zn-doped Bi:2212 samples ($T_c^{max} = 84$ $K$
\cite{AnukoolPhD} - data not shown here).

\begin{figure}[hbtp]
\begin{center}
\includegraphics[width=75mm,keepaspectratio=true]{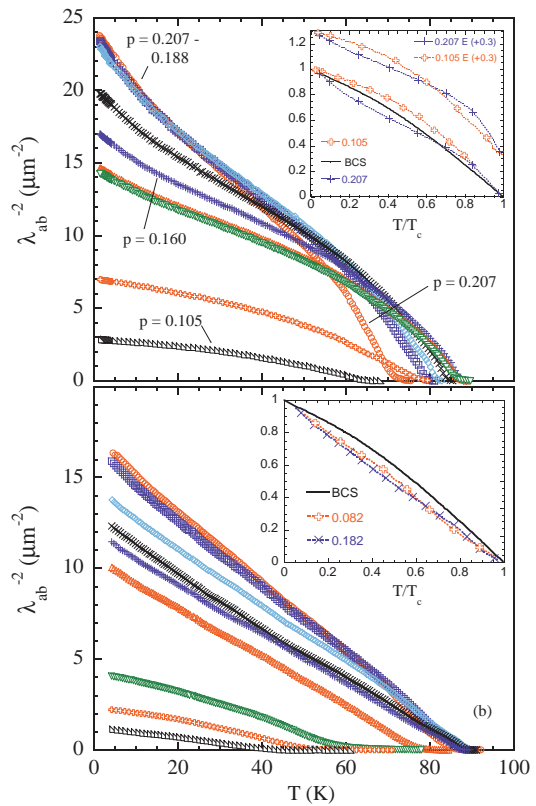}\caption{Color online: the superfluid density
($n_{s}\propto 1/\lambda_{ab}^{2}$) plotted as a function of $T$ for (a) pure Bi:2212
and (b) 15\% Y-doped Bi:2212. Values of $p$ are the same as in Fig. 2. Comparisons
with weak coupling $d$-wave behavior for the highest and lowest values of $p$  are
shown in the insets.  Results assuming  3:3:1 ellipsoids (E) with H $\|$ to the short
side, instead of spheres, are also shown in the upper inset. Here $\lambda_{ab}(0)$
values are factors of 1.23 and 1.1 larger than for spheres for $p$ = 0.207 and 0.105
respectively. } \label{Fig.3}
\end{center}
\end{figure}

\section{Analysis and Discussion}
The data in Figs.~\ref{ch05-fig02}a and~\ref{ch05-fig02}b have been analyzed in the
usual way by summing London's expression $m/m_{max} = 1 - 3\frac{\lambda}{a}\coth
\frac{a}{\lambda} + 3\frac{\lambda^{2}} {a^{2}}$  for  a superconducting sphere of
radius $a$ over the measured particle size distribution and varying $\lambda$ until
$m/m_{max}$ equalled the measured value. The resulting values of $\lambda_{ab}$ are
plotted as $1/\lambda_{ab}^2$ \textit{vs.} $T$ in Figs.~\ref{Fig.3}a and~\ref{Fig.3}b.
For Bi(Y):2212, $1/\lambda_{ab}^2$  \textit{vs.} $T$ is linear from low $T$ up to
$T_c$ for $p$ = 0.12 to 0.183 while for pure Bi:2212 there is evidence for
``super-linear'' behavior. Most of the data are slightly more linear than expected for
a weak coupling BCS $d$-wave state, as shown in the insets to Fig. \ref{Fig.3}.  A
recent compilation based on various techniques \cite{HufnerRev} suggests that
$2\Delta_{max}(0)$ is usually $\sim 5 k_BT_c$ rather than the weak coupling value
$4.28 k_BT_c$ \cite{Won&Maki}.  So we would  not expect large deviations from weak
coupling $d$-wave, but, if anything, larger gaps near the nodes would cause a slower
decrease in $1/\lambda_{ab}^2$ with $T$ at low $T$. The $T$-dependence  for Hg:1223
\cite{Panagop-1223} is close to that for weak coupling $d$-wave, but a slightly
faster, more linear decrease was observed for YBa$_2$Cu$_3$O$_7$ \cite{Panagop-Zn} and
more recently for heavily under-doped YBa$_2$Cu$_3$O$_{6+x}$ crystals \cite{Broun}.
Relatively few over-doped materials have been studied  so the ``super-linear''
behavior could be a general property of clean over-doped cuprates. Note that
$\lambda_{ab}(T)$ has only been reported for optimally-doped Bi:2212 crystals
\cite{Lee, Jacobs} and in Fig. 3a the ``super-linear'' behavior is less marked in this
case, i.~e.~for $p$ = 0.16. However SEM pictures \cite{AnukoolPhD} of the sintered
samples studied here showed that the crystallites of pure Bi:2212 were especially thin
and plate-like, $\sim$ 0.3 $\mu m$ thick, while those of Bi(Y):2212 were much thicker.
It possible that this is could be partly responsible for the ``super-linear''
behavior, as indicated in the inset to Fig. 3a, where  results for  an analysis based
on ellipsoids with 3:3:1 aspect ratios are shown. So although the ``super-linear''
dependence  is potentially an important result it needs to be confirmed by other
measurements.

\begin{figure}[htbp]
\begin{center}
\includegraphics[width=7.5cm,keepaspectratio=true]{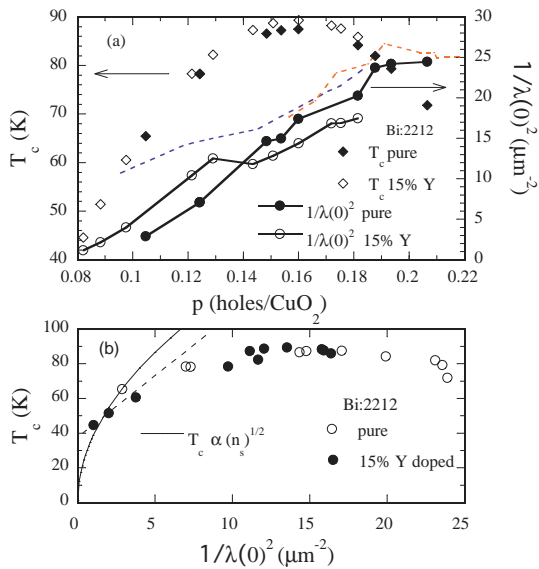}
\caption{ Color on-line: (a) $T_c$ and  $\lambda_{\circ}^{-2}$  are plotted vs. $p$.
The  red and blue dashed lines show data from heat capacity work
\cite{Tallon-paradigm}. (b) Plot of $T_c$ vs. $\lambda(0)^{-2}$, i.e.
$n_s(T\rightarrow 0)$. The dashed line shows a linear variation  while the solid line
shows $T_c$~$\alpha$~$\sqrt{n_s(0)}$ \cite{Broun}. } \label{Lambdap}
\end{center}
\end{figure}
Generally speaking $1/\lambda^{2}$ is related to the low energy quasi-particle weight.
If this were preferentially distributed near the nodes of the $d$-wave superconductor
in $k$-space then $1/\lambda^{2}$ would indeed rise at low temperature as observed.
Alternatively we note that the two chain cuprate, YBa$_2$Cu$_4$O$_8$ shows a similar
``super-linear'' behavior which was ascribed to superconductivity induced in the
chains by the proximity effect at low $T$ \cite{Panagop-124}. This raises the
intriguing possibility that the pairing interaction in Bi:2212 is very $k$-dependent
and that for over-doped Bi:2212 samples, proximity coupling could be playing a role
even for states within the Cu-O$_2$ planes.

   In Fig.~\ref{Lambdap}a data for both samples show a strong
linear increase  in $1/\lambda_{ab}^{2}(0)$ with $p$ for $p \lesssim 0.19$ but
$1/\lambda_{ab}^{2}(0)$ suddenly becomes constant for $p \gtrsim 0.19$. Bearing in
mind that $1/\lambda_{ab}^{2}(0)$ is a measure of the superfluid density $n_s(0)$ as
$T\rightarrow 0$ this strongly supports scenario B for $T^*(p)$.  As shown in
Fig.~\ref{Lambdap}b there is a region where $T_c$ increases linearly with $n_s(0)$,
but the large intercept at $T_c$ = 40 $K$ means that the
 empirical Uemura relation, $T_c$ $\alpha$ $n_s(0)$, \cite{Uemura} does not apply here. The solid line shows that at low $p$ our data are
more compatible with $T_c$  $\alpha$ $\sqrt{n_s(0)}$ found recently
 for YBa$_2$Cu$_3$O$_{6+x}$ \cite{Broun} although for our Bi:2212 samples
  the values of ${n_s(0)}$ are much smaller for  similar values of $T_c$. The plot in Fig.~4b is qualitatively
 consistent with the PG model of Ref. \onlinecite{Tallon-paradigm} but
 the initial increase of $T_c$ with $n_s(0)$ in our data seems to be too sudden to give
 quantitative agreement.  As also shown in Fig.~4a the our direct measurements of $\lambda_{ab}(0)$
 are  consistent with earlier results \cite{Tallon-paradigm} from a mean-field, Ginzburg-Landau (GL)  analysis
 of the field-dependent heat capacity, except at low  $p$ where we find smaller values
 of $n_s(0)$. This is probably because of the large error bars in $n_s(0)$ obtained from the
 heat capacity at low $p$ where the superconducting contribution is small\cite{LoramPriv}.
   In Fig.~4a, the deviation of the two Bi(Y):2212 points at $p$ = 0.121 and 0.129
   from the
 general trend of $1/\lambda_{ab}^{2}(0)$ with $p$ might be connected
with a  1/8th plateau effect~\cite{Koike-01} in $T_{c}$. However simple modification
of the $T_{c}(p)$ line near $p$ = 0.125 will not account for these two anomalous
points.

    Finally we briefly compare our results for $n_s(0)$ with recent STS
 data.\cite {KohsakaDavis}  $n_s(0)$ is a Fermi surface property and if the  density
 of states $N(E)$ in the normal state is strongly energy ($E$) dependent it is
 expected  \cite{Tallon-paradigm} to be given by:
 \begin{equation}
  n_s(0) = \mu_0e^2<v_x^{2}N(E)>
 \end{equation}
 where $v_x$ is the projection of the Fermi velocity along the supercurrent direction
 and the average is taken over an energy range of the order of the superconducting
 gap.  In a BCS-like $d$-wave situation this energy
 range will be $E_F \pm 3\Delta({\textbf{k}})$ where the product of the BCS parameters $u_{\textbf{k}}$ and $v_{\textbf{k}}$
 is finite.  The STS work suggests that for heavily
 underdoped Bi:2212 samples with $T_c$ values of  20 and 45 $K$ there are Fermi arcs (strictly
 speaking in the superconducting state these are  Bogoliubov arcs) whose length in
\textbf{k}-space is still $\sim 1/3$ of that of the large hole-like Fermi surface seen
in
 overdoped samples where there is no pseudogap.   As the Mott insulating state is approached
both techniques show a loss in density of states or spectral weight near $E_F$.  But
our results suggest that $n_s(0)$ is reduced by a factor of 25 when $T_c$ = 40 $K$
while at first sight the STS work only gives a factor of 3 reduction.  Our data can be
reconciled with  STS  if this loss of states  also occurs in the region of the arcs,
but over an energy range smaller than $E_F \pm 2-3 \Delta(\textbf{k})$, so that the
Bogoliubov quasi-particle peaks are still visible in STS.  This is qualitatively
consistent with the model in Ref. \onlinecite{Tallon-paradigm} in which the PG is
usually smaller than the superconducting gap. However we note that a microscopic
theory of this effect also needs to consider what happens when a supercurrent is
produced by uniformly displacing the Fermi surface in \textbf{k}-space.  Assuming that
the PG is not displaced, then this continuity requirement seems to imply that only
regions with a finite density of states at, and close to, $E_F$ will contribute to the
supercurrent.

\section{Summary and Conclusions}
 We report direct evidence that at low $T$ the  superfluid density  of
Bi:2212 falls rapidly for $p\lesssim 0.19$,  which supports scenario B for the
pseudogap in this widely-studied \cite{HufnerRev} compound. For both samples
$1/\lambda_{ab}^{2}(0)$ is extremely small in the heavily under-doped region, down by
a factor $\approx 25$ while $T_{c}$ remains relatively high ($T_{c}\simeq$ 40 $K$). In
conjunction with recent STS work \cite{KohsakaDavis} this could be a useful constraint
for theoretical models. For many values of $p$ there is evidence for a linear
$T$-dependence of the superfluid density over an unusually wide range of $T$ and
preliminary evidence for ``super-linear'' behavior below $20$ $K$ in  over-doped
samples of pure Bi:2212.

\section{Acknowledgements}
 We acknowledge funding from the
EPSRC (U.K.), for the experimental facilities in Cambridge, and further financial
support from The Royal Society, EURYI, and MEXT-CT-2006-039047. W. A. was supported by
the Development and Promotion of Science and Technology Talents Project (D.P.S.T.),
Thailand.  We thank J. W. Loram, S. H. Naqib, J. L. Tallon and E. M. Tunnicliffe for
helpful discussions.


\begin{thebibliography}{99}
\bibitem{Footnote1} To prevent water vapor condensing on the samples after the
quench, the gold sample container was placed in a glass beaker and warmed to room
temperature by a strong stream of flowing gas. This procedure also worked well for the
powder samples that were contained in a gold foil envelope but typically 1 - 2 $mg$
were lost each time.  All samples were periodically overdoped and re-measured to check
that there had been no preferential loss of smaller or larger particles during the
quench.
\bibitem{Footnote2} The sequence is composed of 4
treatments in O$_2$ from 450 to $550^{\circ}C$, then $450^{\circ}C$ in O$_2$ (5) and
again at step 8. Treatments 6, 7, 9 and 10 correspond to various atmospheres,
including vacuum, at temperatures up to $600^{\circ}C$, while 11 represents slow
cooling from 450 to $100^{\circ}C$ in O$_2$. After the initial heating to
$550^{\circ}C$ (steps 1-4) the data lie on the same reversible line. When referred to
the oxygen content derived from the weight changes (top $x$-axis) this line has a
slope of $0.93\pm 0.09$.
\bibitem{Footnote3}  A
commercial (Lake Shore Model DRC-91CA) susceptometer and a homemade one with miniature
coils of 2.6 $mm$ internal diameter were used. The former was convenient for absolute
magnitude and a wider range of AC fields while the latter could be used down to 1.3
$K$ and gave smoother temperature dependences (absorption of paramagnetic oxygen can
give spurious anomalies in magnetic susceptibility measurements between 40 and 60
$K$). They were calibrated by measuring pure lead (Pb) spheres at low enough
frequencies (3.3 or 33.3 $Hz$) to have negligible eddy current signals in the normal
state.   The volume ($V$) of the powder was found from the weight and X-ray density
(6.68 mg/mm$^3$) and hence the signal ($m_{max}$) corresponding to an assembly of
perfectly diamagnetic spheres could be found. In our experience the particles settle
in the sample capsule so loosely that the magnetic interaction between grains is
negligible, but in any case the demagnetization factor of the
 particles in their holder (the bottom of a gelatin  capsule) is approximately 1/3,
  so the local field acting
 on any grain will be very close to the applied field~\cite{Porch-01}.
\bibitem{Footnote4} The London equation $\nabla^{2}B = B/\lambda^{2}$ is linear, so any
 diamagnetism arising from current loops that flow out of the
$ab$-plane is additive. Since there are two in-plane axes, the appropriate averaging
factor for out-of-plane currents and randomly oriented crystallites is  2/3. So for
$\lambda_{c}\gg r$ there is an extra diamagnetic signal of
$\frac{2}{3}\frac{3}{2}\frac{1}{4\pi}\frac{1}{15}r^{2}/\lambda_{c}^{2}$ $emu/cm^{3}$
in the measured susceptibility. With $r =1$ $\mu m$ and $\lambda_{c} =70$ $\mu m$
\cite{CooperForro}, this is only $3\times 10^{-6}$ $emu/cm^{3}$ on the plots of
Fig.~\ref{ch05-fig02}. The effect would be negligible even if $\lambda_{c}$ were a
factor of 10 smaller. In our experience \cite{Cooper-01} $\lambda_{c}/\lambda_{ab}$
often follows the behavior of $\sqrt{\rho_{c}/\rho_{ab}}$ near room temperature, and
this only decreases
 by a factor of 2 between $p = 0.16$ and $p = 0.20$ ~\cite{Forro-01}.
\end{thebibliography}
\end{document}